%% file: vignali.tex
\documentclass[useAMS]{mn2e}
\usepackage{graphicx}
\usepackage{times,mathptm}
\newcommand{\ltsima}{$\; \buildrel < \over \sim \;$}
\newcommand{\simlt}{\lower.5ex\hbox{\ltsima}}
\newcommand{\gtsima}{$\; \buildrel > \over \sim \;$}
\newcommand{\simgt}{\lower.5ex\hbox{\gtsima}}

\newcommand{\cgs}{ ${\rm erg~cm}^{-2}~{\rm s}^{-1}$} 
\newcommand{\lum}{\rm erg~s$^{-1}$}

\def\lesssim{\mathrel{\hbox{\rlap{\hbox{\lower4pt\hbox{$\sim$}}}\hbox{$<$}}}}
\def\gtrsim{\mathrel{\hbox{\rlap{\hbox{\lower4pt\hbox{$\sim$}}}\hbox{$>$}}}}

\def\arcdeg{\hbox{$^\circ$}}
\def\arcmin{\hbox{$^\prime$}}
\def\arcsec{\hbox{$^{\prime\prime}$}}

\def\ab1450{$AB_{1450(1+z)}$}

\def\xray{\hbox{X-ray}}

\def\oiii{\hbox{[O\ {\sc iii}}]}

\def\nev{\hbox{[Ne\ {\sc v}}]}

\def\09104{IRAS~09104$+$4109}
\def\I09104{I09104}

\def\msun{M$_{\odot}$}
\def\edd_ratio{$\log\ L_{\rm bol}/L_{\rm Edd}$}

\def\Nh{{N$_{\rm H}$}}
\def\l58{{$(\lambda L_{\lambda})_{\mbox{{\rm \scriptsize 5.8\micron}}}$}}
\def\lmir2{{$(\lambda L_{\lambda})_{\mbox{{\rm \scriptsize 12.3\micron}}}$}}
\def\s1{{S$_{\mbox{{\rm \scriptsize 3.6\micron}}}$}}
\def\irac2{{S$_{\mbox{{\rm \scriptsize 4.5\micron}}}$}}
\def\f3{{S$_{\mbox{{\rm \scriptsize 5.8\micron}}}$}}
\def\mic8{{S$_{\mbox{{\rm \scriptsize 8\micron}}}$}}
\def\f24{{F$_{\mbox{{\rm \scriptsize 24\micron}}}$}}

\def\asca{{\it ASCA\/}}
\def\chandra{{\it Chandra\/}}

\def\heao1{{\it HEAO-1\/}}

\def\iras{{\it IRAS\/}}
\def\iso{{\it ISO\/}}
\def\spitzer{{\it Spitzer\/}}
\def\scuba{{\it SCUBA\/}}

\def\rosat{{\it ROSAT\/}}

\def\sax{{\it BeppoSAX\/}}
\def\nustar{{\it NuSTAR\/}}

\def\xmm{{XMM-{\it Newton\/}}}
\def\suzaku{{\it Suzaku\/}}

\def\swift{{\it Swift\/}}
\def\integral{{\it Integral\/}}
\def\aj{AJ}

\def\apj{ApJ}
\def\apjl{ApJ}
\def\apjs{ApJS}

\def\aap{A\&A}

\def\mnras{MNRAS}

\def\pasj{PASJ}

\def\nat{Nature}

\def\gca{Geochim.~Cosmochim.~Acta}
 
\def\memsai{Mem.~Soc.~Astron.~Italiana}

\def\procspie{Proc.~SPIE}

\title[An X-ray and mid-IR view of \09104]
{On the nature of the absorber in \09104: the X-ray and mid-infrared view}
\author[C. Vignali et al.]
{
C. Vignali,$^{1,2}$\thanks{E-mail: cristian.vignali@unibo.it.}
E. Piconcelli,$^{3}$ 
G. Lanzuisi,$^{4,5}$
A. Feltre,$^{6}$  
C. Feruglio,$^{7}$
R. Maiolino,$^{3}$
\newauthor 
F. Fiore,$^{3}$ 
J. Fritz,$^{8}$ 
V. La Parola,$^{9}$ 
M. Mignoli$^{2}$ and 
F. Pozzi$^{1}$ \\ \\
$^{1}$ Dipartimento di Astronomia, Universit\`a degli Studi di Bologna, 
Via Ranzani 1, 40127 Bologna, Italy \\
$^{2}$ INAF -- Osservatorio Astronomico di Bologna, Via Ranzani 1, 
40127 Bologna, Italy \\
$^{3}$ INAF--Osservatorio Astronomico di Roma, Via di Frascati 33, 
00040 Monteporzio Catone, Italy \\
$^{4}$ INAF--IASF Roma, Via Fosso del Cavaliere 100, 00133 Roma, Italy \\
$^{5}$ INAF--IASF Bologna, Via Gobetti 101, I-40129 Bologna, Italy \\
$^{6}$ Dipartimento di Astronomia, Vicolo Osservatorio 3, 35122 Padova, 
Italy \\
$^{7}$ Institut de RadioAstronomie Millimetrique, 300 rue de la Piscine, 
Domaine Universitaire, 38406 Saint Martin d'Heres, France \\
$^{8}$ Sterrenkundig Observatorium, Universiteit Gent, Krijgslaan 281 S9, 
B-9000 Gent, Belgium \\
$^{9}$ INAF--IASF Palermo, Via Ugo La Malfa 153, 90146 Palermo, Italy
}

\begin{document}

\date{Accepted 2011  June 2.  Received 2011 May 31; in original form 2011 January 7}

\volume{416}
\pagerange{2068--2077} \pubyear{2011}

\maketitle

\label{firstpage}

\begin{abstract} 
We present a long ($\approx76$~ks) \chandra\ observation of \09104, a 
hyper-luminous galaxy, optically classified as a Type~2 AGN hosted in 
a cD galaxy in a cluster at $z=0.442$. We also report on the results 
obtained by fitting its broad-band spectral energy distribution. The 
Compton-thick nature of this source (which has been often referred 
to as an `archetype'' of Compton-thick Type~2 quasars) was formerly 
claimed on the basis of its marginal detection in the PDS instrument 
onboard \sax, being then disputed using \xmm\ data.  Both \chandra\ 
analysis and optical/mid-IR spectral fitting are consistent with the 
presence of heavy ($\approx1-5\times10^{23}$~cm$^{-2}$), but not 
extreme (Compton-thick) obscuration. 
However, using the mid-IR and the \oiii\ emission as proxies of the 
nuclear hard \xray\ luminosity suggests the presence of heavier 
obscuration. The 54-month \swift\ BAT map 
shows excess hard \xray\ emission likely related to a nearby 
($z$=0.009) Type~2 AGN, close enough to \09104\ to significantly 
enhance and contaminate its emission in the early \sax\ PDS data. 
\end{abstract}


\section{Introduction}
\label{intro}
\09104\ (hereafter referred to as \I09104) is a hyper-luminous 
infrared cD galaxy at $z$=0.442 residing in the 
core of a rich cluster of galaxies (Kleinmann et al. 1988). Its
optical spectrum is characterized by the presence of narrow emission
lines, although broad Balmer and MgII lines were observed in polarized
light (Hines \& Wills 1993; Tran, Cohen \& Villar-Martin 2000). These
remarkable results, based on optical observations, suggested the
presence of a powerful quasar hidden in \I09104. Subsequent \xray\
observations confirmed this ``picture''. While \sax\ data below 10~keV
were ``contaminated'' by the intra-cluster emission, the weak signal
detected by the PDS instrument in the range \hbox{$\approx$~15--60~keV} (in a
34~ks exposure) was interpreted by Franceschini and collaborators
(2000, hereafter F00) as the primary emission of the buried quasar,
emerging at very high energies because of Compton-thick absorption
(i.e., above $\approx10^{24}$~cm$^{-2}$; in the case of \I09104, the
estimated lower limit on the column density was
$5\times10^{24}$~cm$^{-2}$). The detection of a strong iron $K\alpha$
emission line, with equivalent width (EW) of $\approx$1--2~keV --
consistent with a reflection scenario for the quasar emission in the
2--10~keV energy range -- further supported the Compton-thick nature 
of \I09104\ in F00, and similar conclusions, though reached with
limited counting statistics (but much higher spatial resolution), were
drawn from a 9.1~ks \chandra\ observation (Iwasawa, Fabian \& Ettori
2001, hereafter I01).

Doubts on these conclusions were recently cast by Piconcelli et al. (2007, 
hereafter P07) on the basis of the analysis of a 14~ks \xmm\ observation. 
In particular, although both a transmission- and a 
reflection-dominated model provided a good fit to the quasar emission, the 
iron line EW was lower ($\approx400$~eV) than expected in the case of a 
Compton-thick scenario (EW$\approx1-2$~keV), but it was higher 
than model predictions for a few$\times10^{23}$~cm$^{-2}$ obscuration 
(Ghisellini, Haardt \& Matt 1994).
Furthermore, the source bolometric luminosity was well recovered using 
the \hbox{2--10~keV} flux obtained in a transmission model (after 
removal of the significant contribution from the cluster, see P07) and 
assuming a reasonable \xray\ bolometric correction ($\S$4 in P07). 
According to these authors, the weak detection in the PDS instrument 
onboard \sax\ was explained by the presence of a heavily obscured, possibly 
Compton-thick galaxy (NGC~2782; Zhang et al. 2006) in the large 
(1.3\arcdeg\ FWHM) field of view of the PDS. 

To date, the number of ``certified'' (i.e., secured by good-quality 
\xray\ spectra and, possibly, high-energy detection) Compton-thick 
Active Galactic Nuclei (AGN) is still limited, being settled around 
\hbox{$\approx40-50$} and mostly at low ($\simlt0.05$) redshifts (see 
Comastri 2004 for a review on this topic, and Della Ceca et al. 2008 
for un updated list of Compton-thick AGN). Large numbers of 
Compton-thick AGN candidates at higher redshifts have been recently 
claimed, but their identification as heavily obscured AGN is often 
derived from other, indirect arguments [e.g., optical emission-line 
strength, combined mid-infrared (mid-IR) and optical criteria, stacked 
\xray\ emission; e.g., Alexander et al. 2008, 2011; Fiore et al. 2008, 2009; 
Lanzuisi et al. 2009; Bauer et al. 2010; Gilli et al. 2010, and references 
therein; Vignali et al. 2010] than direct, good-quality \xray\ spectroscopy. 

Clearly, the lack of broad-band \xray\ instruments within the same 
observatory (as in \sax) is currently limiting the detection and 
spectral characterization of Compton-thick AGN; in this regard, the 
\integral-IBIS (Beckmann et al. 2009) and \swift\ BAT (Tueller et 
al. 2008) surveys have brought to the detection of additional very 
local Compton-thick AGN candidates, but their confirmation has always 
required follow-up observations with \chandra, \xmm\ or \suzaku\ 
(e.g., Ueda et al. 2007; Eguchi et al. 2009; Comastri et al. 2010; 
Severgnini et al. 2011; Burlon et al. 2011). To overcome this 
observational problem will necessarily rely on the next-generation 
hard \xray\ imager that will fly on \nustar.

In this context, it is clear the role of our present investigation to 
assess the ``true'' nature of \I09104, i.e., whether it is Compton 
thick or Compton thin. If Compton thick, it would be one of the few 
AGN of this class beyond the local Universe supported by moderately 
good-quality \xray\ spectroscopy (for other moderate/high-redshift 
Compton-thick AGN, see Norman et al. 2002; Iwasawa et al. 2005; 
Comastri et al. 2011; Feruglio et al. 2011; Gilli et al. 2011). 
We tackle this issue through a multi-wavelength approach consisting of 
(a) the analysis of a long ($\approx76$~ks) and recent (2009) \chandra\ 
observation of \I09104\ and (b) the reconstruction of its optical and 
mid- and far-infrared (far-IR) spectral energy distribution (SED) in terms of 
stellar plus reprocessed AGN emission, using the code developed by Fritz, 
Franceschini \& Hatziminaoglou (2006, hereafter F06). 
While (a) will provide further clues on the best-fitting model characterizing 
the \xray\ emission of \I09104, (b) will allow us to estimate the geometry and 
coverage of the obscuring matter as well as its column density, which will be 
compared to that derived from X-rays. 

Finally, we use known correlations found in AGN to estimate the 
expected intrinsic \hbox{2--10}~keV emission. In particular, we use 
the 5.8\micron, 12.3\micron\ and \oiii5007\AA\ luminosities as a proxy 
of the nuclear luminosity following the correlations reported by 
Lanzuisi et al. (2009) and Gandhi et al. (2009) for the mid-IR, and 
Mulchaey et al. (1994) and Heckman et al.  (2005) for the 
optical. Furthermore, we also compute the \hbox{2--10}~keV over 
\nev3426\AA\ flux ratio to possibly place further constraints on the 
amount of absorption. 

A $\Lambda$-cosmology with \hbox{$H_{0}$=70~km~s$^{-1}$~Mpc$^{-1}$}, 
\hbox{$\Omega_{\rm M}$=0.3} and \hbox{$\Omega_{\Lambda}$=0.7} 
(Spergel et al. 2003) is adopted. 

\section{Chandra data}

\subsection{X-ray data analysis}
\label{xray_analysis}
\I09104\ was observed on Jan 6$^{th}$ 2009 for 76.16~ks with the 
Advanced CCD Imaging Spectrometer (ACIS; Garmire et al. 2003) and the 
I3 CCD at the aimpoint. The observation (OBS\_ID=10445) was carried 
out in very faint mode.  Since the source is in a massive cooling-flow 
cluster (Fabian \& Crawford 1995), showing a significant radial 
temperature gradient, the background subtraction is highly critical, 
despite the brightness of \I09104.  Therefore, to extract spectra, we 
decided to adopt a strategy similar to that applied by I01 to the 
previous 9.1~ks \chandra\ observation, dating back to 1999. We 
extracted the source spectrum from a 1\arcsec-radius region (large 
enough to include \hbox{$\approx$~85--90}~per~cent of the on-axis 
\chandra\ ACIS PSF) and the background from the surrounding 
1\arcsec--2\arcsec\ annulus, to possibly account for a significant 
fraction of the cool-core cluster thermal emission. 
The source extraction region was chosen from the 
\xray\ image above 2~keV, where the contrast between \I09104\ and the 
extended emission is maximized. At softer energies, the source does not 
appear to be point-like, and its location seems consistent with the bright 
core of the cluster emission. 
Because of its superior PSF, using \chandra\ data allows us to better 
evaluate the contribution due to the thermal emission -- 
whose detailed analysis is beyond the purposes of this work -- 
than in the \xmm\ observation (P07). 
We note that after background removal, $\approx$~700 source counts are 
available for spectral fitting. 
The resulting spectrum has been grouped with 
a minimum of 20 counts per bin to apply the $\chi^{2}$ statistic and fitted 
with {\sc xspec v12.5.0} (Arnaud 1996). 

The results obtained from this source spectral extraction has been checked 
- as reported in the following - using a different strategy, 
consisting of fitting the background contained in the  1\arcsec--2\arcsec\ 
annulus, freezing its spectral parameters to their best-fitting values, and 
then applying this background model (properly scaled to account for the 
factor 3 difference in area between source and background) to the source 
spectral components (see $\S$\ref{xray_fitting}). 

Hereafter, errors are quoted at the 90~per~cent confidence level for 
one interesting parameter (i.e., $\Delta\chi^{2}=2.71$; Avni 1976); 
all spectral fits include absorption due to the line-of-sight Galactic column 
density of \hbox{$N_{\rm H}=1.42\times10^{20}$~cm$^{-2}$} 
(Kalberla et al. 2005), and Anders \& Grevesse (1989) abundances are assumed. 

\subsection{X-ray spectral fitting}
\label{xray_fitting}
To reproduce the \chandra\ data over the 0.5--7~keV energy range, we
have adopted two models, similarly to the fitting strategy applied by
P07 to \xmm\ data. The first model (hereafter referred to as 
``transmission'' model) comprises an absorbed powerlaw component plus 
an iron emission line.  The second model (referred to as 
``reflection'' model) includes a reflection component ({\sc pexrav} in 
{\sc xspec}) and an iron emission feature. These models are reported 
as {\sf A} and {\sf B} in Table~\ref{xray_spectrum}, where {\sf A2} 
differs from {\sf A1} because the powerlaw photon index is fixed to 
1.8, as expected and typically found in case of AGN emission (e.g., 
Piconcelli et al. 2005, and references therein). 
%
\input{vignali.tab1.tex}
%
Furthermore, both models include a soft component due to the residual 
(i.e., not totally accounted for by background subtraction) cluster thermal 
emission. 
It is also possible that there might be a contribution to soft X-rays 
from photoionized gas at the inner nucleus, as suggested by I01; 
this hypothesis would be consistent with the hard-to-soft \xray\ flux 
ratio ($\approx$~4~per~cent) observed in \I09104, which is comparable 
to that typically measured in Sey~2 galaxies (e.g., Matt et al. 2004; 
Piconcelli et al. 2008, and references therein). However, a detailed 
analysis of the soft \xray\ emission, limited by the low number of 
counts below 2~keV ($\approx200$) and the lack of deep \xray\ grating 
data, is beyond the purposes of this paper, since the main goal of our 
work consists of a proper characterization of the hard ($>$2~keV) 
emission. 

In the upper part of Table~\ref{xray_spectrum}, we report the \xray\ 
spectral results for both models using a ``standard'' background subtraction 
within {\sc xspec}, where the source$+$background spectrum is extracted from 
a circular region of radius 1\arcsec\ centered on the position of \I09104, 
and the background spectrum from a surrounding annulus, as anticipated in 
$\S$\ref{xray_analysis}. 
The lower part of Table~\ref{xray_spectrum} reports the same models using 
a different background subtraction strategy; at first, the background, using 
its own response matrices, is fitted with a thermal component 
plus a low-normalization powerlaw component. 
The best-fitting parameters are then frozen, 
and the source$+$background spectrum is fitted using these components 
(multiplied by a normalization constant fixed to 0.33, i.e., 
the source/background area ratio) and those ascribed to the nucleus of \I09104. 
As evident from Table~\ref{xray_spectrum} and already reported above, 
there is still need for a thermal component, possibly due to the 
residuals at soft \xray\ energies obtained from the best-fitting of 
the background spectrum. By using the latter background subtraction 
procedure, errors on the parameters are typically lower. 

Whatever the method adopted to account for background-subtraction is, 
the transmission model is preferred in terms of quality of the fit 
(see column (7) in Table~\ref{xray_spectrum}). In this model, the 
column density is \Nh$\approx(1-5)\times10^{23}$~cm$^{-2}$, closely resembling 
the absorption values found by P07. The iron line, compatible with being 
neutral or mildly ionized, has an EW$\approx300-400$~eV (in the rest frame of 
the source), is consistent, within errors, with the value found by P07 
(and similarly for the line flux) and with 
that obtained by I01 using lower exposure \chandra\ data. 
We note that these large EW values, though not extreme, are slightly higher 
than expected in the case of fluorescence iron line produced by the 
obscuring matter (Ghisellini et al. 1994)\footnote{The predicted 
EW for the iron line depends on the inclination angle of the torus, 
although this dependence is not strong given the column density range derived 
from \xray\ spectral fitting of \I09104\ data.} but appear consistent with 
previous \xmm\ and \suzaku\ observations of Compton-thin Seyfert~2 galaxies 
(e.g., Guainazzi, Matt \& Perola 2005; Fukazawa et al. 2011). 
Given the large stellar mass estimated via SED fitting 
(which is described in $\S$\ref{sed}), 
$M_{\star}\approx4.8\times10^{11}$~\msun, and the 
known mass-metallicity relation (Lequeux et al. 1979; Savaglio et al. 2005), 
we may expect some metal (more specifically, iron) enhancement in \I09104, 
hence an increased strength of the Fe emission line, although it is not 
straightforward to quantify such effect. 
For instance, NGC~4388, a local Sey~2 galaxy with a column density 
comparable to that of \I09104, shows a similarly large iron K$\alpha$ EW 
(Iwasawa et al. 2003; see also Beckmann et al. 2004 and Shirai et al. 
2008 for significant EW differences due to continuum variability), although 
it is not hosted by a massive galaxy.
On the other hand, we note that Mrk~231 shows a relatively low EW for the 
iron K$\alpha$ line, despite being a Compton-thick AGN (Braito et al. 2004).
More convincing hints for a Compton-thin absorption towards \I09104\ are 
likely provided by the high-energy constraints (see $\S$\ref{bat}). 

\begin{figure}
\includegraphics[angle=-90,width=0.47\textwidth]{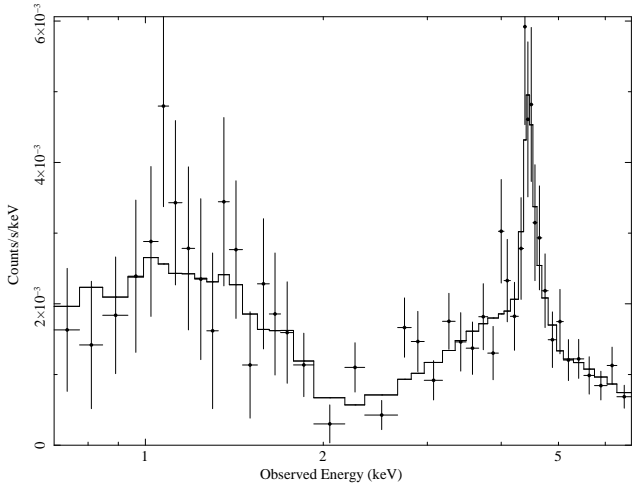}
\caption{
\chandra\ datapoints of \I09104\ fitted with an absorbed powerlaw 
plus an iron line (transmission model; {\sf A1} model in the top part of 
Table~\ref{xray_spectrum}) and thermal emission, ascribed to the cluster.}
\label{chandra_spectrum}
\end{figure}

The 90~per~cent upper limit on the EW for He-like and H-like iron line at 
6.7~keV and 6.97~keV is $\approx$~85~eV. 
Although the powerlaw is characterized by an apparently flat photon index, 
the solution with a more physically acceptable $\Gamma=1.8$ still provides 
an adequate fit to the data (see Table~\ref{xray_spectrum}). 
The source spectrum, fitted using a transmission model (including a neutral 
iron line), is shown in Fig.~\ref{chandra_spectrum}. 
%

The 2--10~keV flux\footnote{Since the currently used version of 
{\sc psextract}, the tool adopted to extract the \xray\ spectrum from 
\chandra\ data, does not take into account the correction for the 
encircled energy fraction enclosed in the source extraction region, we 
compute average ``correction factors'' to be applied to the fluxes 
using the \chandra\ Proposer's Guide. At the on-axis position of our 
target, these factors, included in the flux and luminosity values 
reported hereafter, are $\approx$~10 and $\approx$18~per~cent for the 
0.5--2~keV and 2--10~keV bands, respectively.} is 
$\approx(3.8-4.2)\times10^{-13}$~\cgs, corresponding to an intrinsic 
(i.e., absorption-corrected in the case of models {\sf A1/A2}) AGN 
luminosity (placing the contribution due to the residual thermal 
emission to 0) of $\approx(2.7-3.1)\times10^{44}$~\lum\ (where the 
range of values is ascribed to the differences in the method adopted 
for background subtraction). 

Although the low-energy (below $\approx$~6~keV) range provided by 
\chandra\ is apparently not well suited for placing significant 
constraints on the reflection component, we modeled the data 
using a Compton reflection component from neutral matter ({\sc 
pexrav} model in {\sc xspec}; Magdziarz \& Zdziarski 1995); we adopted 
cos i=0.9 as inclination angle for the reflection component, 
which is equivalent to assuming a rather face-on geometry for the 
reflecting material, likely to be associated with the inner walls of 
the obscuring torus; we chose a cut-off energy of 200~keV for the 
primary continuum, and metal abundances were fixed to the solar value. 
We also fixed the photon index of the continuum to 1.8 to facilitate 
the spectral fitting with the reflection model over the limited energy 
bandpass provided by \chandra\ and get a reasonable constraint on the 
strength of such component. 
%
%
The \chandra\ data require a reflection-dominated spectrum, where the 
primary powerlaw normalization is formally consistent with 0. 
From a statistical point of view, the 
reflection model provides a worse description of the \xray\ data, 
regardless of the adopted method to subtract the background (see 
Table~\ref{xray_spectrum}).  
The iron $K\alpha$ emission line EW is $\approx400-500$~eV, which is 
not easily reconcilable with what expected in a typical reflection 
scenario. 

We note that the photon index derived by our spectral fittings in case of 
a transmission model ({\sc A1}) is flat. This result can suggest that an 
additional reflection component may be present, causing the observed 
\xray\ photon index to be flatter than expected from AGN emission 
(e.g., Piconcelli et al. 2005); however, a ``mixed'' absorption and 
reflection model is not able to provide any improvement to the 
previous spectral solutions, calling for higher quality data over a 
possibly more extended energy interval.

Overall, it seems that any residual cluster emission has been taken 
into account properly by our procedure of background removal and 
fitting, thanks to the good on-axis PSF provided by \chandra.  For 
instance, the ionized iron line components related to extended 
(cluster) emission may be at the origin of the large iron line EW 
previously reported (F00). 

The \chandra\ data presented in this paper provide consistent 
results with those published by P07; although the photon statistics is 
higher in the \xmm\ observation (which also guarantees data over a broader 
energy range), \chandra\ data allow us for a better treatment of the 
contamination from the cluster in virtue of its better PSF. We find 
that $\approx$~90~per~cent of the flux is ascribed to the active 
nucleus over the 2--10~keV range, to be compared with 
\hbox{$\approx$~30--35}~per~cent derived by P07, where the source counts were 
extracted from a much larger [37\arcsec\ (pn) -- 40\arcsec-radius 
(MOS)] region.

\section{High-energy constraints from the 54-month Swift BAT data}
\label{bat}
We used the 54-month \swift\ BAT data, processed using the {\sc 
Bat\_Imager} software (Segreto et al. 2010), to constrain the 
high-energy emission of \I09104. The source is not detected in any of 
the BAT sub-bands, the 3-$\sigma$ upper limits to the source \xray\ 
emission (derived using a Crab-like spectrum) being 
$\approx1.9\times10^{-12}$, $\approx3.3\times10^{-12}$, and 
$\approx4.8\times10^{-12}$~\cgs\ in the 15--30~keV, 15--70~keV, and 
15--150~keV bands, respectively. 
The lowest energy upper limit provided by BAT is consistent, within 
uncertainties, with the flux derived in the 20--30~keV energy range by 
F00 using \sax\ data ($\approx2.5\times10^{-12}$~\cgs). The flux 
limit in the BAT map at the position of \I09104\ (but see below for 
concerns about possible contamination) is too loose to place any 
constrain on the two \xray\ models presented in 
$\S$\ref{xray_fitting}.

\begin{figure}
\includegraphics[angle=0,width=0.47\textwidth]{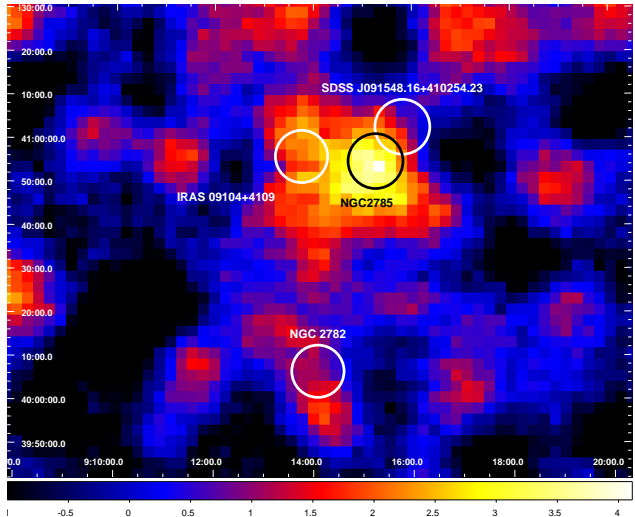}
\caption{
\swift\ BAT 54-month significance map (15--30~keV band) of the area 
around \I09104, clearly showing hard \xray\ emission close but likely not 
associated to the $z=0.442$ quasar. 
The position of NGC~2782, previously claimed to contaminate the PDS flux 
of \I09104\ (P07), is also shown, along with the positions of NGC~2785 
(black circle) and SDSS~J091548.16$+$410254.23; see text for details. 
All circles have radii of 6\arcmin, broadly corresponding to the 
positional uncertainty provided by BAT for weak sources. 
The color scale indicates the intensity level of the pixels in the map.}
\label{bat_map}
\end{figure}

As shown in Fig.~\ref{bat_map}, there is ``excess'' emission (at the 
$\approx4\sigma$ level, i.e., just below the detection threshold 
adopted for BAT maps) close (at $\approx$~17\arcmin) to our source of 
interest, although such emission, at first glance, appears related to 
another object. It is reasonable to believe that the upper limit to 
the hard \xray\ emission of \I09104\ reported above is actually 
contaminated, at least in part, by the nearby source's PSF wings and, 
as such, must be intended as very conservative and poorly indicative 
of the effective high-energy emission of \I09104. This finding 
provides support to the hypothesis of a Compton-thin absorber towards 
\I09104.

Using the NASA Extragalactic Database (NED) and literature, we have 
attempted to assign a known counterpart to the source closest to 
\I09104.  Taking into account the positional uncertainty in the BAT map 
($\approx$~6\arcmin, corresponding to the size of the circles in 
Fig.~\ref{bat_map}), at the position of the BAT hard \xray\ emission 
we found NGC~2785, a disky galaxy at $z=0.009$, whose Sloan Digital 
Sky Survey (SDSS) spectrum is characterized by a very red continuum 
and a large H$\alpha$/H$\beta$ ratio ($\approx20.9$ at face value, or 
$\approx12.5$ taking into account the correction for the average 
host galaxy absorption component), 
being highly suggestive of a heavily extincted object. 
The ``corrected'' H$\alpha$/H$\beta$ ratio is consistent with the 
value reported by Veilleux et al. (1995) using a different optical spectrum. 
Under reasonable assumptions, this line flux ratio translates into a 
E(B$-$V)$\approx$1.5--1.9. 
Furthermore, the H$\alpha$/[NII] and H$\beta$/[OIII] line ratios are 
consistent with this source being a Type~2 AGN. The association of 
the BAT excess with this source appears as the most likely 
possibility, as also suggested by their good spatial coincidence 
(see Fig.~\ref{bat_map}). 
%
%
At present, the only \xray\ constraint to the emission of 
NGC~2785 below 10~keV (over a broader energy range than that allowed 
by the shallow \rosat\ All Sky Survey, RASS, where the source is not 
detected) can be derived by an \asca\ GIS observation, where the 
source, located at a large off-axis angle from the nominal pointing, 
is not detected. Assuming a simple powerlaw model with $\Gamma$=1.8, 
the derived 3$\sigma$ upper limit to its 2--10~keV flux is 
$\approx1.2\times10^{-13}$~\cgs. Using the de-reddened \oiii\ emission 
to estimate the intrinsic 2--10~keV flux (adopting the correlation 
found by Mulchaey et al. 1994), we derive an 
\xray\ flux of $\approx2.9\times10^{-11}$~\cgs. At face value, the 
difference between the upper limit to the hard \xray\ flux derived 
from \asca\ data and that estimated from the \oiii\ intensity can be 
explained assuming an absorber with a column density of at least 
$\approx2.6\times10^{24}$~cm$^{-2}$. However, the upper limits to the \xray\ 
emission of NGC~2785 above 10~keV derived from the BAT instrument are 
well below the extrapolation of the absorbed powerlaw, thus casting 
some doubts on either the \xray\ flux derived from the \oiii\ 
measurement or the adopted model (or both). Pointed and sensitive 
\xray\ observations of NGC~2785 are required to shed light on all 
these issues. 

At a close position, we found another possible (but less likely) 
optical counterpart, a broad-line SDSS quasar at \hbox{$z=2.32$}, 
SDSS~J091548.16$+$410254.23.  This quasar has a tentative detection 
($\approx2.4\sigma$) in the RASS at $\approx$~19\arcsec. However, from 
the BAT map, it seems clear that this source does not provide a 
significant contribution to the observed signal. 

Although the association of the high-energy emission with NGC~2785 
requires further support, we note that BAT data have clearly shown 
that contamination in the early PDS data is a valuable hypothesis. As 
a consequence, serious doubts are cast on the Compton-thick nature of 
\I09104, since it was mostly motivated by the admittedly marginal 
($\approx2.5\sigma$) detection in the PDS instrument and the large 
equivalent width of the iron K$\alpha$ line (F00), which most probably 
suffered from the contribution of ionized iron features associated 
to the thermal emission in \sax\ data. 

Furthermore, from the BAT map it seems clear that NGC~2782 
($z=0.008$), the Seyfert~2 galaxy thought to contaminate the hard 
\xray\ emission of \I09104\ in the PDS (P07), has marginal emission in 
the hard X-rays (see Fig.~\ref{bat_map}), and its distance from the 
$z=0.442$ quasar ($\approx50$\arcmin) is beyond any possible 
contamination in the BAT map.

\section{Spectral energy distribution: multi-wavelength data and fitting}
\label{sed}
To provide a better understanding of the matter responsible for the \xray\ 
absorption towards \I09104, we have investigated the source broad-band 
properties using the data available in literature. The scientific goal 
consists of using the reprocessed emission to constrain the 
physical/geometrical parameters of the absorber (the ``postulated'' torus 
in AGN unification schemes) and estimate the bolometric luminosity of the 
AGN due to accretion of matter onto the black hole. 
This study requires that all the possible source components emitting in 
the UV, optical, near-IR, mid-IR, and far-IR bands are properly taken into 
account in terms of stellar (galaxy) plus nuclear (AGN) components. 
The broad-band analysis has been carried out by using most of the 
photometric datapoints available in Table~B.4 of Ruiz et al. 
(2010). In particular, in our SED fitting analysis, we used the \xmm\ 
Optical Monitor (OM) data (filters UVW2 and UVW1), the SDSS magnitudes 
(excluding the $i$-band magnitude, given its contamination from strong \oiii\ 
emission), the published \spitzer\ IRAC data at 3.6\micron\ and 5.8\micron, 
and MIPS data at 24\micron. The source was also detected by \iras\ at all 
wavelengths\footnote{For the flux densities at the \iras\ wavelengths, 
we refer to the association of \I09104\ with the \iras\ Faint Source 
Catalog object F09105$+$4108.} 
up to 60\micron, while at longer wavelengths (\iras\ at 
100\micron, \scuba\ at 450\micron\ and 850\micron), only upper limits are 
available. 
We remind the reader of the non-simultaneity for most of the data used 
in our analysis. This issue, common to most investigations dealing 
with SED fitting, likely plays a minor role, given the long timescales 
for variability expected in case of reprocessed emission by dust. 
We also note that our photometric points include both nuclear (i.e., 
AGN-related) and galaxy emission, with both components being accounted 
for by the adopted broad-band SED-fitting procedure, as described in the 
following. 
Finally, as {\em a-posteriori} check, we have verified 
that the broad-band modeling adopted to reproduce the emission from \I09104\ 
is consistent, in the mid-IR, with the \spitzer\ IRS spectrum available in 
literature (discussed in $\S$\ref{irs_spectrum}). 

Data have been modeled using the code developed by F06, which assumes 
that the dust, responsible for the reprocessing of the nuclear 
radiation, consists of graphite and silicate grains distributed in a 
``flared disc'', ``smooth'' geometry. Recent high-resolution, 
interferometric mid-IR observations of nearby AGN (e.g., Jaffe et 
al. 2004; Tristram et al. 2007; Meisenheimer et al. 2007; 
Tristram \& Schartmann 2011) have confirmed the presence of 
a geometrically thick, torus-like dust distribution on pc-scales; 
however, data strongly suggest that this torus is irregular or ``clumpy''.
\xray\ variability on short time-scales provides further support to such 
clumpy-absorption scenario (Risaliti et al. 2007, 2010). In this
regard, many codes have been developed over the last decade to deal 
with clumpy dust distributions (e.g., Nenkova, Ivezi{\'c} \& Elitzur 
2002; Nenkova et al. 2008a,b, N08a,b hereafter; Schartmann et 
al. 2008; H{\"o}nig \& Kishimoto 2010). 
We note, however, that fitting broad-band SED data, based on sparse 
photometric datapoints, with either smooth or clumpy model is not 
supposed to provide significantly different results (see Dullemond \& 
van Bemmel 2005) in terms of derived AGN bolometric luminosity. 
In the following, we will mainly focus on the best-fitting parameters 
obtained by adopting the smooth dust distribution (F06); however, a 
discussion based on the broad-band data fitting using the clumpy code 
described by N08a,b will also be provided.

The SED fitting is based on a multi-component analysis, which has been 
adopted often in recent literature for the analysis of obscured AGN (e.g., 
Vignali et al. 2009; Hatziminaoglou, Fritz \& Jarrett 2009; Pozzi et 
al. 2010). 
The observed UV to far-IR SED of \I09104\ has been de-composed in three 
distinct components: stars, having the bulk of the emission in the 
optical/near-IR; hot dust, mainly heated by UV/optical emission due to gas 
accreting onto the supermassive black hole and whose emission peaks 
somewhere between a few and a few tens of microns; cold dust, principally 
heated by star formation. 

The stellar component has been included using a set of Simple Stellar 
Population (SSP) spectra of solar metallicity and ages ranging from 
$\approx 1$~Myr to $\approx$7.3~Gyr, which corresponds to the time 
elapsed between $z$=4 (the redshift assumed for the stars to form) 
and $z=0.442$ in the adopted cosmology. 
A common value of extinction is applied to stars of all ages, and 
a Calzetti et al. (2000) attenuation law has been adopted (R$_V=4.05$). 
Assuming a higher redshift for the event of star formation does not 
provide significantly different results, while a Milky Way extinction curve 
produces a slightly worse fit than the one reported here 
(see Table~\ref{smooth_sed_fitting}). 

\input{vignali.tab2.tex}

The AGN emission, dominating in the mid-IR range, comprises a thermal 
component due to reprocessing of the nuclear AGN emission, a 
scattering component and a disc component. In particular, we used an 
extended version of the model grid described by F06. 

Given the presence of data above 24\micron, also an emission component 
coming from colder, diffuse dust, likely heated by star-formation 
processes, has been included in the fitting procedure, and it is 
represented by templates of known starburst galaxies (see Vignali et 
al. 2009 for details). 

Similarly to Vignali et al. (2009) - to which we refer for further details on 
the adopted procedure and its limitations - we evaluate the goodness of the 
fit using a ``merit function'', defined 
as $MF=\sum_{i=1}^{N_O}\left(\frac{M_i-O_i}{\sigma_i}\right)^2/N_O$, where 
$N_O$ is the number of observed datapoints, $M_i$ and $O_i$ are the model 
and observed flux densities, respectively, of the $i-th$ photometric band, 
and $\sigma_i$ is the corresponding observed error. 

\begin{figure}
\includegraphics[angle=-90,width=0.5\textwidth]{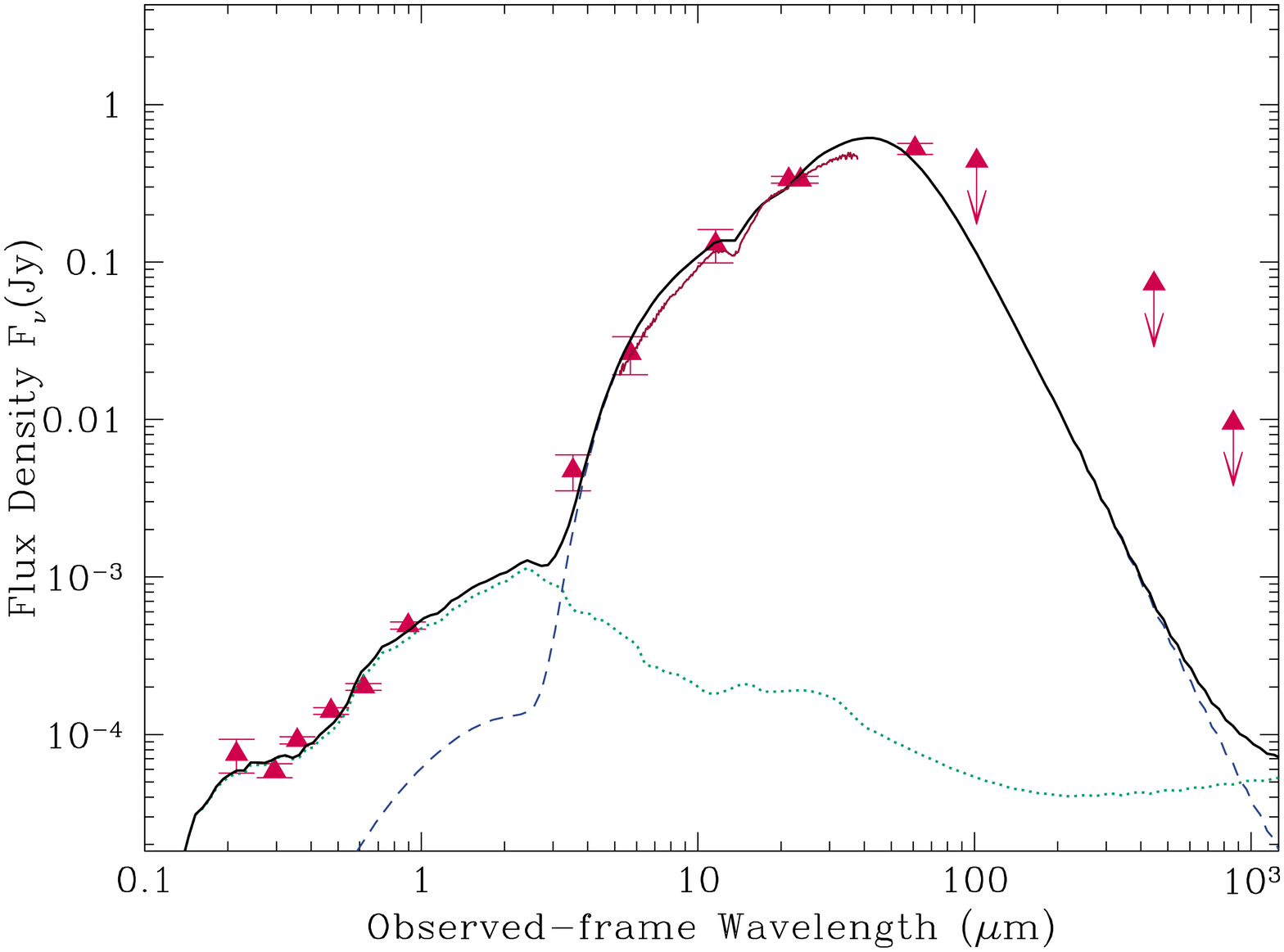}
\caption{
Observed-frame SED of \I09104, from the \xmm\ Optical Monitor (OM) 
ultraviolet datapoints to the \scuba\ 850\micron\ datapoint. 
3-$\sigma$ upper limits refer to non-detections (\iras\ 100\micron\ 
and both \scuba\ wavelengths). The total SED (thick black line) is 
the summed contribution of the stellar (green dotted line) and AGN 
components (blue shaded line). The continuous red line over the 
wavelength range \hbox{$\approx5-38$\micron} represents the IRS 
spectrum, without any normalization applied. We note that the 
best-fitting solution obtained using the F06 model is consistent with 
the IRS spectrum in terms of normalization and reproduction of the 
9.7\micron\ silicate absorption feature.}
\label{sed_calzetti_bestfitting_solar_met}
\end{figure}

From the SED fitting, there is no need for invoking cold dust ascribed 
to starburst emission, as suggested by the absence of detected emission in 
the far-IR above 60\micron\ (Fig.~\ref{sed_calzetti_bestfitting_solar_met}). 
The upper limits at longer wavelengths appear fully consistent with AGN 
emission (see also Rowan-Robinson 2000 for similar conclusions). 
In the UV/optical band, the stellar component has a moderate extinction 
[E(B-V)=0.16]. 

The best-fitting solution (Fig.~\ref{sed_calzetti_bestfitting_solar_met}) 
in terms of MF (see Table~\ref{smooth_sed_fitting}) provides a torus 
full covering angle of 140\arcdeg, corresponding to a covering factor of 
$\approx$~90~per~cent. 
The half opening angle, 20\arcdeg\ ((180-140/2)), is larger than 
the one suggested by the presence of a well collimated radio jet 
(Kleinmann et al. 1988; Hines \& Wills 1993) and a narrow optical 
ionization cone (Crawford \& Vanderriest 1996); see Fig.~3 of 
Taniguchi et al. (1997). We note, however, that none of the torus 
models in the grid used in this paper has a covering angle larger than 
140\arcdeg; to limit the degeneracies in the model parameters, only 
covering angles of 60, 100, and 140 degrees are actually used (see 
Fritz et al. 2006). These values refer to three possible ``classes'' 
of geometric solutions for the torus: a disk-like torus, a 
moderate-coverage torus, and one covering most of the nuclear source. 
Our best-fitting SED delineates a configuration where a significant 
fraction of the innermost regions of \I09104\ is hidden by the torus; 
in these terms, its covering angle can be considered consistent with 
indications obtained with previous observations. 

The extinction at 9.7\micron\ along the equatorial plane\footnote{Since the 
angle between the line of sight and the equatorial plane is zero, the 
extinction along the line of sight and that along the equatorial plane 
coincides. Angles different from 0 would imply an extinction 
in the direction of the observer lower than the equatorial one.} 
is relatively high ($\tau(9.7)$=3) but not 
extreme, as one might expect in case of Compton-thick absorption, 
and corresponds to a column density of $\approx1.1\times10^{23}$~cm$^{-2}$ 
(assuming a Galactic gas-to-dust ratio conversion). 
This can be considered broadly consistent with those derived from \xray\ 
spectral analyses [from both \xmm\ (P07) and \chandra\ data; 
see $\S$\ref{xray_fitting}]. 
Given the discrete values for the torus model grid parameters, it is 
not straightforward to associate uncertainties to the reported values 
but it is possible to evaluate how much solid the best-fitting 
solution is. Analysis of the degeneracies in smooth torus parameters 
(see Pozzi et al. 2010 for details) and corresponding uncertainties 
indicates that the value of the covering angle of the torus 
(140\arcdeg) is well determined, and $\tau(9.7)$ along the equatorial 
plane reaches at most the value of 6 (90~per~cent confidence level), 
which means a doubling of the column density of the absorber. In all 
the solutions provided by F06 code for \I09104, an absorber with a 
corresponding column density of at least $\approx10^{23}$~cm$^{-2}$ 
(but never extreme) is required by the fit. 

Overall, the best-fitting solution for the stellar plus nuclear 
emission of \I09104\ (Table~\ref{smooth_sed_fitting}) obtained using, for the 
reprocessed nuclear component, the smooth dust distribution (F06), is able to 
reproduce most of the observed photometric datapoints as well as the 
shape of the IRS spectrum and the 9.7\micron\ silicate absorption 
feature, as discussed in $\S$\ref{irs_spectrum}. 
The resulting accretion-related bolometric luminosity is 
\hbox{$\approx(2.3-3.8)\times10^{47}$}~\lum\ (where the range provides 
approximately the 90~per~cent uncertainty on the luminosity). 
%

Given the relatively large number of photometric data available and 
the presence of mid-IR spectroscopic coverage for \I09104, this source 
can be considered a good target for a comparison of smooth vs. clumpy 
dust distribution; however, for an exhaustive comparison on a large 
AGN sample, we refer to Feltre et al. (2011). Similarly to the SED 
fitting described above, we used a stellar plus AGN emission to 
account for the broad-band emission of \I09104. In particular, for the 
latter component we used the clumpy dust models from N08a,b; 
according to these models, the dust is distributed in 
clouds, hence there is always a finite probability for 
an unobscured (i.e., direct) view of the AGN, irrespective of the 
viewing angle (e.g., Nikutta, Elitzur \& Lacy 2009). 
Among the latest publicly available grid models (end of September 2010), 
we adopted those which are considered best suited for Type~2 AGN, 
where the direct component is likely minor or negligible (R. Nikutta, 
private communication).  This choice is also motivated by the 
necessity of considering the contribution, at optical and near-IR wavelengths, 
of the host galaxy stellar component, which is visible in optical 
images. Including both stellar and direct nuclear components would 
have provided a over-estimation of the emission at wavelengths 
short-wards a few microns by the model. 

\begin{figure}
\includegraphics[angle=-90,width=0.5\textwidth]{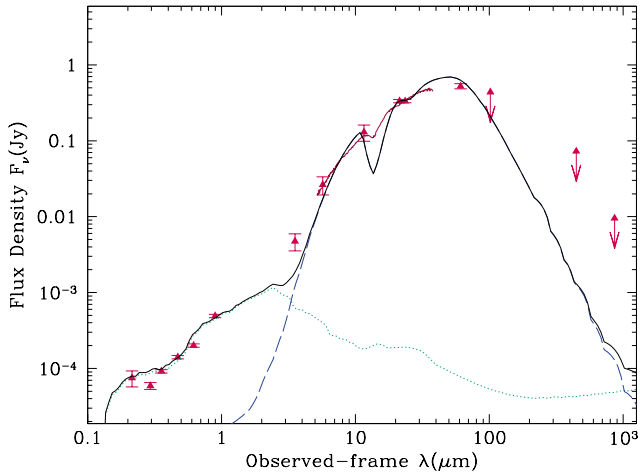}
\caption{
Observed-frame SED of \I09104\ obtained by fitting the same 
photometric points as in 
Fig.~\ref{sed_calzetti_bestfitting_solar_met}. The stellar emission 
(green dotted line) is accounted for using a set of SSPs, while the 
AGN emission is reproduced using a clumpy dust distribution (blue 
shaded line). The total SED is marked by the thick black line, while 
the IRS spectrum is plotted as a continuous red line. The Draine \& 
Lee (1984) optical properties for the silicates have been assumed here 
(see text for details).} 
\label{sed_clumpy_std}
\end{figure}

The clumpy model provides a good fit to the observed data in 
terms of data-to-model residuals (MF$\approx$4.4). However, we note 
two major problems in dealing with this solution (hence, with the 
parameters derived for the clumpy torus): 
(a) the model is not able to reproduce the near-IR (3.6\micron) data, 
irrespective of the assumed stellar population fitting the optical data. As 
extensively discussed by Mor, Netzer \& Elitzur (2009) and, more 
recently, by Deo et al. (2011), this emission, 
not properly accounted for by Nenkova et al. grid models, needs an extra 
hot-dust component (parameterized by a blackbody with temperature of 
$\approx$~1400~K), possibly associated with graphite grains (which, 
given a higher sublimation temperature than silicate grains, can 
survive closer to the black hole; see Mor \& Trakhtenbrot 2011). 
At face value, this result may 
cast some doubts on the parameters derived in the mid-IR in case of 
clumpy models. In fact, the extra model component appears hard to justify, 
since it should be already accounted for by the radiative transfer treatment, 
as in smooth torus models. We note, however, that using the 
Draine \& Lee (1984) optical properties for the silicate component 
(as in Nenkova et al. 2002) instead of the Ossenkopf, Henning \& Mathis (OHM; 
1992), which have been assumed in N08a,b and 
recent works (Nikutta et al. 2009), the discrepancy between model and data 
in the near-IR decreases (see Fig.~\ref{sed_clumpy_std}).  
The second major concern is that (b) the best-fitting clumpy solution 
to the photometric data is not able to reproduce the shallow 9.7\micron\ 
absorption feature due to silicates in IRS spectrum. This result, 
which will be discussed further in $\S$\ref{irs_spectrum}, is likely due 
to the fact that such SED-fitting solution requires the presence of 
many clouds along the equatorial plane. Their combined contribution 
manifests itself in a deep silicate feature, which is actually not 
observed as such in IRS spectrum (see Fig.~\ref{sed_clumpy_std}). 

Since smooth-torus models are able to account, in a self-consistent 
way, all of the major components related to direct, scattered and 
reprocessed AGN emission, we prefer this solution for \I09104. 
However, we caution the reader against over-interpreting our 
simplistic comparison between the smooth and the clumpy solutions. 
Extended and more exhaustive checks need to be carried out 
on a sizable sample of AGN (Feltre et al. 2011) before 
drawing general conclusions on the capability, by either model, to 
account for the reprocessed nuclear emission. 

For what concerns the smooth torus solution, we note that the SED 
fitting presented by Taniguchi et al. (1997) suggested the additional 
presence of cold dust distributed in a much larger outer torus, not 
included in the radiative transfer modeling. This dust layer would 
increase the silicate absorption at 9.7 micron, which has been found 
to be relatively shallow in IRS data. About the other ``inner'' torus 
parameters, their best-fitting solution implies a large covering 
factor (0.98, not too different from our solution; see discussion in 
$\S$\ref{sed}) and a larger inclination with respect to the equatorial 
plane (30\arcdeg) than ours. 

Finally, we may compare the outcome of our SED fitting in terms 
of gas mass with results from literature. Assuming a typical 
dust-to-gas mass ratio of 0.01, the dust mass obtained by our SED 
fitting implies a large gas mass ($\approx5\times10^{10}$~\msun), a 
factor $\approx2-3$ higher than the upper limit derived by Evans et 
al. (1998) through CO measurement. However, both estimates have large 
uncertainties, mainly related to the underlying assumptions (e.g., 
significant variations in the dust-to-gas mass ratio across different 
galaxy populations have been recently reported by Santini et al. 2010). 

\subsection{IRS spectrum of \I09104}
\label{irs_spectrum}
Our own analysis of the \spitzer\ IRS spectrum, covering a wavelength range 
$\approx5-38$\micron\ in the observed frame, confirms the results 
reported by Sargsyan et al. (2008; see also Deo et al. 2009); 
besides the 9.7\micron\ silicate feature in absorption, suggestive of 
obscuration, the IRS spectrum is featureless, except for the very shallow 
silicate absorption feature, and entirely consistent with the 
observed mid-IR datapoints 
(see Fig.~\ref{sed_calzetti_bestfitting_solar_met}). 
The constraints available for the typically most prominent PAH features, 
not observed in \I09104\ (see also Sargsyan et al. 2008), are consistent with 
the lack of a significant starburst component, 
as found from both the SED fitting analysis (see $\S$\ref{sed}) and using IRAS 
(Rowan-Robinson 2000) and \iso\ data (Taniguchi et al. 1997). 
For completeness, we note that in case of a strong radiation field (as 
in the case of our target), the PAHs can hardly survive, if not properly 
screened or produced in regions far away from the active nucleus. 

We may use the IRS spectrum to make some further considerations about 
the smooth vs. clumpy solutions for the matter responsible for the 
reprocessing of the nuclear emission in the mid-IR described above. 
For what concerns the smooth-torus solution, we note that fitting the 
IRS spectrum alone (without taking into account any stellar component, 
because of the IRS wavelength coverage) provides, as 
best-fitting solution, exactly the same found to reproduce the 
source photometry (see Table~\ref{smooth_sed_fitting}). Both the shape 
of the continuum and the 9.7\micron\ silicate absorption feature are 
accounted for reasonably well (see 
Fig.~\ref{sed_calzetti_bestfitting_solar_met}). 
About the clumpy solution, using the Draine \& Lee (1984) optical 
properties for the silicates produces a good fit to the IRS 
continuum and 9.7\micron\ silicate feature which, on the other side, 
is never properly reproduced by clumpy models with OHM silicate dust 
composition. This is opposite to what has recently been reported in 
literature (e.g., Nikutta et al.  2009). Overall, our results on clumpy tori 
applied to \I09104\ indicate that the OHM composition may be rejected, being 
unable to reproduce the near-IR datapoints and the silicate absorption 
feature. Clearly, this result, based on a single source, needs to be 
tested on larger samples. We finally note that in neither case the 
best-fitting clumpy solution to the IRS spectrum is the same one obtained by 
fitting the photometric data.

\section{Discussion}
\label{discussion}
Two of the evidences that brought F00 towards an interpretation of 
\I09104\ as a Compton-thick AGN, i.e., the hard \xray\ emission 
and the large iron K$\alpha$ line EW, have been discussed both by P07 
using the \xmm\ data and in this paper using \chandra\ data. While the 
emission above 10~keV detected by the PDS instrument onboard \sax\ is 
very likely associated to a local, obscured AGN (see $\S$\ref{bat}), 
the large iron emission line EW observed by \sax\ was probably 
originated by a blend of emission features due to the AGN and the 
cluster. The latter emission has been properly accounted for by the 
analysis adopted in this paper thanks to the higher \chandra\ spatial 
resolution with respect to previous \xray\ data which has limited the 
contamination by extended emission (see $\S$\ref{xray_fitting}). 
The analysis of \chandra\ data, coupled with the SED results (under 
the assumption of a smooth dust distribution), suggests that 
Compton-thin absorption is a likely solution for the absorber close to 
\I09104. However, the upper energy boundary covered with sufficient 
statistics by \chandra\ is $\approx$~6~keV (observed frame); 
an extended energy range, 
possibly coupled with simultaneous data above 10~keV, would be useful 
to further support this transmission scenario. 
To provide a broader view of the possible solutions for such absorber, 
in the following we use known relations to estimate the intrinsic 
\hbox{2--10}~keV luminosity of the source (to be compared with the observed 
luminosity), once the cluster emission has been taken into account. 
The following relations are characterized by significant dispersions; 
using them for a single object may suffer from significant 
limitations, therefore any conclusion drawn in the following should be 
treated with care. 

Given the AGN dominance in the mid-IR, the IRS spectrum of \I09104\ is 
used to derive the rest-frame 12.3\micron\ flux density and, adopting 
the Gandhi et al. (2009) correlation (which, obtained from a 
sample of 42 Seyfert galaxies with near diffraction-limited mid-IR 
imaging, links the mid-IR luminosity to the 2--10~keV luminosity; 
see, for a recent application, Vignali et al. 2010), we are able to 
estimate the expected \xray\ emission for this source. On the basis of 
the SED fitting results (see $\S$\ref{sed}), we do not expect a 
significant contribution from star formation at these wavelengths. 

The predicted AGN 2--10~keV luminosity 
\hbox{($\approx8.9\times10^{45}$}~\lum) is higher than the measured value 
(i.e., not corrected for the absorption) in case of a transmission 
scenario \hbox{($\approx1.2-1.3\times10^{44}$~\lum)}. The 
predicted/observed luminosity ratio is large and suggestive of 
Compton-thick obscuration. However, if we use the 5.8\micron\ 
luminosity and the correlation reported in Lanzuisi et al. (2009), 
which has been obtained for a sample of quasars (i.e., at luminosities 
comparable to that of \I09104), albeit of limited size, we find that 
the expected 2--10~keV luminosity is $\approx10^{45}$~\lum, much 
closer to the observed source luminosity but still suggesting heavy 
obscuration. 

The hard \xray\ emission of \I09104\ can be estimated also using 
\oiii-based correlations, under the assumption, motivated by unified 
schemes for AGN, that the \oiii5007\AA\ line is a good proxy of the 
nuclear emission. Using the value for the \oiii\ luminosity of \I09104\ 
taken from Reyes et al. (2008), not corrected for extinction within the 
narrow-line region (NLR), we obtain a 2--10~keV luminosity 
of 5.2$\times10^{45}$~\lum\ and 3.5$\times10^{45}$~\lum\ using the 
correlation, valid for Type~1 AGN, of Mulchaey et al. (1994) and 
Heckman et al. (2005), respectively. 
These values lie in between those obtained from mid-IR estimators and 
discussed above. 

As extensively described in Vignali et al. (2010), these correlations 
may suffer from the still limited number of objects used to derive 
them and, in particular, to the small number of AGN at high 
\xray\ luminosity, i.e., in the quasar regime (indicatively above 
10$^{44}$~\lum, where also \I09104\ lies). For these reasons, 
once more we advise against over-interpreting our estimates of the 
intrinsic nuclear luminosity via the methods reported here, given the 
overall large uncertainties which characterize these correlations and 
their application to a single object. However, these results may 
indicate that some extra absorption with respect to that currently 
measured by \chandra\ might be present in \I09104. 

Finally, we can measure the ratio between the AGN rest-frame 
\hbox{2--10}~keV flux prior to absorption correction and the \nev\ 
emission line (as available from the SDSS spectrum), and compare the 
resulting value ($\approx$~80--90) to the recent findings published by 
Gilli et al. (2010). This ratio is expected to decrease for increasing 
column density, since the \xray\ emission suffers from photo-electric 
absorption, while the \nev\ emission, produced in the NLR, 
can be considered a proxy of the intrinsic nuclear power, 
although, given its rest-frame wavelength and the fact that it is 
likely produced in the inner, probably most obscured part of the NLR, 
it may suffer from extinction. 
While values above few hundreds are definitively an indication 
of lack of obscuration, almost all objects with ratios below 15 are 
Compton thick (see the luminosity plot in Fig.~2 of Gilli et al. 2010 
for objects at comparable redshift). Hence, no firm conclusion on the 
amount of obscuration close to \I09104\ can be inferred from the 
\xray/\nev\ ratio. 
%
%

\section{Summary}
\label{summary}
We have presented sensitive \chandra\ data (i.e., with significantly 
higher photon statistics than those reported by I01) and UV/optical to 
far-IR SED fitting of the narrow-line quasar \I09104\ at $z=0.442$ in 
order to characterize the properties of the matter responsible for the 
\xray\ absorption and reprocessing of the nuclear radiation in the 
mid-IR (i.e., the torus envisaged by AGN unification schemes). The 
source detection in the PDS instrument onboard \sax\ more than ten 
years ago has brought \I09104\ to be classified as a Compton-thick AGN (F00).

In the following, we summarize the main results achieved with the 
multi-wavelength analysis presented in this work. 

\begin{description}
\item[$\bullet$]
The \xray\ spectrum is in favour of a Compton-thin hypothesis, which 
has also been suggested by P07 from the analysis of \xmm\ data. 
\item[$\bullet$]
The 54-month \swift\ BAT map in the 15--30~keV energy range has 
clearly shown that excess hard \xray\ emission, if present (since 
formally below the BAT detection threshold adopted for source 
catalogs), is likely to be ascribed to NGC~2785, a nearby ($z=0.009$) 
Type~2 AGN at $\approx$~17\arcmin\ from \I09104. The properties of its 
SDSS spectrum are suggestive of heavy extinction. 
\item[$\bullet$]
Further support to the Compton-thin hypothesis comes from observations 
at longer wavelengths, where the obscuring matter (torus), responsible 
for all of the observed emission up to 60\micron, can be well 
parameterized, providing a $\tau(9.7)$=3.0, consistent with the 
absorption measured through X-rays (using a Galactic gas-to-dust ratio 
conversion). The covering factor of the absorber is 
$\approx$~90~per~cent. Broad-band photometric data are well 
reproduced by a stellar component plus a smooth dust distribution (as 
described by F06); the silicate absorption feature at 9.7\micron\ 
observed in the \spitzer\ IRS spectrum is also properly accounted for 
by the same torus model.  On the other hand, clumpy-torus models 
(N08a,b), which likely provide a more realistic 
(physical) description of the obscuring matter (e.g., Tristram et 
al. 2007), show problems in reproducing the source photometry and 
spectroscopy at the same time. 
\item[$\bullet$]
Using the mid-IR (at 12.3\micron\ and 5.8\micron, derived directly 
from the IRS spectrum) and the \oiii\ emission to predict the hard 
\xray\ luminosity of \I09104\ provides some space for the presence of 
additional absorption to be added to the value of column density derived 
from \chandra\ data.
\end{description}

The scientific case of \I09104\ strongly supports the importance of 
broad-band, possibly simultaneous \xray\ spectral coverage to define 
the source complexities properly and provide a robust estimate of the 
amount of absorption towards the source. In this regard, hard 
\xray\ spectroscopic and imaging capabilities, as those provided by 
\nustar\ (planned to be launched in 2012), might provide the final answer 
on the nature of the absorber in \I09104. 

\section*{Acknowledgments}
The authors thanks the referee for his useful comments that improved
the quality of the paper, F. Panessa for providing
\spitzer\ information about \I09104\ before publication, and A. Comastri, 
R. Gilli, R. Nikutta, P. Ranalli, P. Severgnini, D. Vergani, G. Zamorani for 
insightful suggestions. Partial support from the Italian Space Agency 
(contracts ASI--INAF I/023/05/0, ASI I/088/06/0, and 
ASI/INAF/I/009/10/0) is acknowledged. 
This research has made use of the NASA/IPAC Extragalactic Database 
(NED) which is operated by the Jet Propulsion Laboratory, California 
Institute of Technology, under contract with the National Aeronautics 
and Space Administration, and of data obtained from the Chandra Data Archive 
and software provided by the Chandra X-ray Center (CXC).

\end{document}

%% file: vignali.tab1.tex
\begin{table*}
\centering
\caption{Best-fitting spectral parameters for the \chandra\ spectrum of \I09104.}
\label{xray_spectrum}
\footnotesize
\begin{tabular}{ccccccc}
\hline
Model & $kT$ & $\Gamma$ & \Nh & E$_{K\alpha}$ & EW$_{K\alpha}$ & $\chi^2$/dof \\
(1)   & (2)  &    (3)   & (4) &      (5)      &        (6)     &     (7)        \\
\hline
\multicolumn{7}{c}{\sf Source: R=1\arcsec; Back: annulus 1\arcsec--2\arcsec} \\ \\
{\sf A1} & 3.74$^{+5.73}_{-1.24}$ & 0.70$^{+0.71}_{-0.40}$ & 2.59$^{+1.42}_{-1.14}$ & 6.41$\pm{0.06}$        & 425$^{+164}_{-140}$ & 30.0/39  \\
{\sf A2} & 4.07$^{+9.16}_{-1.58}$ & 1.8 (fr)               & 4.29$^{+1.25}_{-0.82}$ & 6.41$\pm{0.06}$        & 316$^{+210}_{-115}$ & 35.1/40  \\
{\sf B}  & 1.44$^{+1.00}_{-0.52}$ & 1.8 (fr)               &                        & 6.42$\pm{0.06}$        & 530$^{+182}_{-184}$ & 53.2/40  \\
\hline
\multicolumn{7}{c}{\sf Source: R=1\arcsec; Back: annulus 1\arcsec--2\arcsec\ -- Separate source and back fitting} \\ \\
{\sf A1} & 3.81$^{+4.38}_{-1.20}$ & 0.74$^{+0.44}_{-0.26}$ & 2.81$^{+0.61}_{-0.72}$ & 6.42$\pm{0.06}$        & 337$^{+149}_{-112}$ & 26.4/39  \\
{\sf A2} & 4.02$^{+4.81}_{-1.38}$ & 1.8 (fr)               & 4.49$^{+0.73}_{-0.69}$ & 6.41$^{+0.07}_{-0.05}$ & 274$\pm{118}$       & 30.8/40  \\
{\sf B } & 1.44$^{+0.65}_{-0.75}$ & 1.8 (fr)               &                        & 6.43$\pm{0.06}$        & 423$^{+136}_{-147}$ & 52.0/40  \\
\hline
\hline
\end{tabular}
\begin{minipage}[l]{10.2cm}
Notes --- The source counts are extracted from a circular region 
centered on \I09104\ of radius 1\arcsec, and the background from the 
surrounding annulus of size 1\arcsec--2\arcsec. 
The top part of the table refers to the ``standard'' case where the 
spectral parameters are obtained by subtracting the background to the 
source$+$background spectrum within {\sc xspec}. 
The bottom part of the table reports the same spectral parameters when a 
different approach (fit to the background counts and subsequent fit to 
the source data, once the background components, properly normalized by the 
source area/background area, are frozen to their best-fitting values) 
is adopted. 
(1) \xray\ modeling: A1: absorbed power-law model $+$ iron line (transmission 
model); A2: the same, with $\Gamma$ frozen to 1.8, as expected for AGN 
emission; B: reflection $+$ iron line (reflection model), 
with $\Gamma$ frozen to 1.8. 
In both models, a thermal component has been included; 
(2) temperature of the gas in the cluster (fitted with a {\sc mekal} model 
within {\sc xspec}; 
(3) power-law photon index; 
(4) column density (in units of 10$^{23}$~cm$^{-2}$); 
(5) rest-frame energy of the iron K$\alpha$ line (in keV), whose width 
is fixed to 10~eV;  
(6) rest-frame equivalent width of the iron line (in eV); 
(7) $\chi^2$/degrees of freedom (dof).
\end{minipage}
\end{table*}

%% file: vignali.tab2.tex
\begin{table}
\centering
\caption{Best-fitting SED parameters using a ``smooth'' torus model.}
\label{smooth_sed_fitting}
\footnotesize
\begin{tabular}{ccccc}
\hline
$\theta$ & $\tau_{9.7\micron}$ & Cov. Angle & $E(B-V)$ & MF  \\ 
    (1)  &           (2)      &      (3)   &    (4)   & (5) \\
\hline
0 & 3.0 & 140 & 0.16 & 4.9 \\
\hline
\hline
\end{tabular}
\begin{minipage}[l]{0.48\textwidth}
Notes --- For a full description of the parameters, see Fritz et al. (2006). 
(1) Angle between the line of sight and the equatorial plane of the torus 
(in degrees); 
(2) optical depth at 9.7~\micron; 
(3) full covering angle of the torus (in degrees); 
(4) $E(B-V)$ of the stellar component;  
(5) ``merit function'' (sort of $\chi^{2}$), 
used to evaluate the goodness of the fit (see text for details).
\end{minipage}
\end{table}